# Observation of Higher-Order Nodal-Line Semimetal in Phononic Crystals

Qiyun Ma,[1] Zhenhang Pu,[1] Liping Ye,[1] Jiuyang Lu,[1] Xueqin Huang,[2] Manzhu Ke,[1] Hailong He,[1, *] Weiyin Deng,[1,*] and Zhengyou Liu[1,3,*]

[1]Key Laboratory of Artificial Micro- and Nano-structures of Ministry of Education and School of Physics and Technology, Wuhan University, Wuhan 430072, China

[2]School of Physics and Optoelectronics, South China University of Technology, Guangzhou 510640, China.

[3]Institute for Advanced Studies, Wuhan University, Wuhan 430072, China

[*]Corresponding authors: hehailong@whu.edu.cn; dengwy@whu.edu.cn; zyliu@whu.edu.cn

Higher-order topological insulators and semimetals, which generalize the conventional bulk-boundary correspondence, have attracted extensive research interest. Among them, higher-order Weyl semimetals feature two-fold linear crossing points in three-dimensional (3D) momentum space, 2D Fermi-arc surface states, and 1D hinge states. Higher-order nodal-point semimetals possessing Weyl points or Dirac points have been implemented. However, higher-order nodal-line or nodal-surface semimetals remain to be further explored in experiments in spite of many previous theoretical efforts. In this work, we realize a second-order nodal-line semimetal in 3D phononic crystals. The bulk nodal lines, 2D drumhead surface states guaranteed by Zak phases, and 1D flat hinge states attributed to $k_z$-dependent quadrupole moments, are observed in simulations and experiments. Our findings of nondispersive surface and hinge states may promote applications in acoustic sensing and energy harvesting.

Since the discovery of topological insulators, topological phase of matter has been a thriving research field [1-3]. Different from topological insulators with bulk gaps, topological semimetals are exotic topological phases characterized by nontrivial bulk band crossings [4,5]. The known Weyl [6-8] or Dirac [9,10] semimetals, nodal-line semimetals (NLSMs) [11,12], and nodal-surface semimetals [13] host the degeneracies at zero-dimensional (0D) points, 1D lines, and 2D surfaces, respectively. A crucial feature of these intriguing topological phases is that the bulk topological property supports gapless boundary states, known as the conventional bulk-boundary correspondence. These topological boundary states live in $(d-1)$-D boundaries for a $d$-D system regardless of whether their bulks are gap or gapless [14]. Particularly, NLSMs possess 1D nodal lines that can form fruitful topological configurations, such as nodal rings [15,16], nodal chains [17,18], nodal links [19,20], nodal knots [21,22], and nodal nets [23]. The nodal lines usually support a nontrivial $\pi$ Zak phase, leading to the drumhead surface states [5]. Notably, recent investigations revealed that nodal lines can related with non-Abelian topology described by quaternion charge, and exhibit braiding topological structures [24-29].

Higher-order topological phases generalize the conventional bulk-boundary correspondence, which are reflected by the topological boundary states localized at boundaries of co-dimension $d_c > 1$ [30-33]. Higher-order topological insulators, including second order and third order, have been implemented in phononic crystals (PCs) [34-39], photonic crystals [40-47] and electric circuits [48,49]. Naturally, higher-order topology can emerge in topological semimetal, giving rise to the concept of higher-order topological semimetal [33]. Higher-order Weyl semimetals, which feature 3D bulk Weyl points, non-closed 2D surface Fermi arcs and 1D hinge arcs terminated at the projection of Weyl points, were proposed [33,50,51] and soon afterwards realized [52-54]. At the same time, the intriguing bulk-hinge correspondence in higher-order Dirac semimetals [55] has also been demonstrated in PCs [56,57] and photonic crystals [58]. More recently, higher-order topological semimetals with 1D nodal lines have been proposed [53,59-63], which exhibit 2D drumhead surface states and 1D hinge states simultaneously. These hinge states attribute to nontrivial quadrupole moments [53,59], Wannier centers [60], or second Stiefel-Whitney topological charges [61,62]. Despite many previous efforts, it is highly desirable to explore higher-order NLSMs in experiment.

In this work, taking the advantages of PC with ease of construction and measurement, we theoretically design and experimentally realize a 3D acoustic second-order NLSM. In this NLSM, 2D drumhead surface states on the $x$-$y$ surface are guaranteed by Zak phases distributed in the $k_x$-$k_y$ plane, while 1D flat hinge states on the hinges between the $x$-$z$ surface and $y$-$z$ surface attribute to $k_z$-dependent quadrupole moments. We first illustrate the first-order and second-order topological properties of the NLSM in a tight-binding model, then design and fabricate PC samples to observe the bulk nodal lines, drumhead surface states, and flat hinge states. The experimental results, together with the theoretical and simulated ones, confirm the existence of the second-order NLSM.

We start from a tight-binding model with simple cubic structure that involves stacking a 2D square lattice along the $z$ direction, as presented in Fig. 1(a). Four sites A-D of a unit cell are connected by intracell hoppings $t_1$ (golden cylinders) and intercell hoppings $t_0$ (silver cylinders) in each layer. The interlayer couplings with same magnitudes to $t_1$ and $t_0$ are chiral and give rise to gauge flux in the reduced 2D lattices. The Bloch Hamiltonian reads

$$H = \begin{pmatrix} 0 & 0 & h_{13} & h_{14} \\ 0 & 0 & h_{23} & h_{24} \\ h_{13}^* & h_{23}^* & 0 & 0 \\ h_{14}^* & h_{24}^* & 0 & 0 \end{pmatrix},$$

with $h_{13} = t_1 + t_0 e^{ik_x}$, $h_{14} = t_1 + t_0 e^{ik_y}$, $h_{23} = (t_1 + t_0 e^{-ik_y})e^{-ik_z}$, and $h_{24} = t_1 + t_0 e^{-ik_x}$, where $\mathbf{k} = (k_x, k_y, k_z)$ is the Bloch wavevector. The lattice constants in all directions are set unitary for convenience. Figure 1(b) shows the distribution of nodal line in the first Brillouin zone (BZ). And Fig. 1(c) depicts the bulk dispersions along the high-symmetry lines, in which the red line denotes the nodal-line dispersion between the second and third bands. The nodal-line degeneracy is protected by the mirror symmetry along the Γ-M direction at $k_z = 0$ plane combined with chiral symmetry (see Ref. [64]). Due to the existence of chiral symmetry, the nodal-line degeneracy (red line) in Γ-M line is strictly confined to zero energy. In addition, a nodal-surface degeneracy emerges at $k_z = \pi$ (or $-\pi$) plane between the first (third) and second (fourth) bands, protected by $S_2$ screw symmetry.

The NLSM has first-order topological property described by Zak phase. Considering $k_x$ and $k_y$ as parameters, the Zak phase of the model for fixed $k_x$ and $k_y$ can be obtained by using the Wilson loop method [65,66] along the $k_z$ direction across the first BZ. The distribution of Zak phase in the $k_x$-$k_y$ plane is shown in the upper panel of Fig. 1(d). The surface BZ [blue plane in Fig. 1(b)] is divided into two square regions (gray and green) by the projections of nodal lines (red dashed lines). The Zak phase in the green region is $\pi$, signifying the existence of drumhead surface states on the $x$- $y$ surface for these $k_x$ and $k_y$ [67]. The projected surface dispersion confirms our prediction, as shown in the lower panel of Fig. 1(d). The gray color represents the bulk states closest to zero energy, and the green plane denotes the drumhead surface states that correspond to $\pi$ Zak phase. The NLSM hosts 2D drumhead surface states, exhibiting the conventional bulk-boundary correspondence.

Besides the first-order topology, the NLSM also has second-order topological property described by the $k_z$-dependent quadrupole moment, thus is a second-order NLSM. Specifically, for a fixed $k_z$, the model can be considered as a 2D square lattice with a synthetic gauge field that is introduced by interlayer hopping. The nonzero gauge field induces half-quantized quadrupole moments, as shown in Fig. 1(e). The $k_z$-dependent quadrupole moment $q_{xy}$ is determined by using the nested Wilson loop method [30]. Due to the existence of nodal-line degeneracy, bulk quadrupole moment is ill-defined at $k_z = 0$. The nontrivial quadrupole moment implies the existence of hinge states. This can be verified by Fig. 1(f) which depicts the hinge dispersions (orange lines) along the $k_z$ direction. The hinge dispersions appearing at whole 1D projected BZ highlight the bulk-hinge correspondence. Unlike the topological drumhead surface states on the $x$-$y$ surface, the surface states (green grids) on the $x$-$z$ and $y$-$z$ surfaces are topologically trivial.

We implement the second-order NLSM in the PC, with a photograph shown in Fig. 2(a). The unit cell is illustrated in Fig. 2(b), which contains four acoustic resonators connected by oblique cylinder tubes. The resonator is spiraling upward with a displacement of $a/4$ where $a = 36.0$ mm is the lattice constant, to make the strengths of interlayer and intralayer couplings equal, corresponding to the tight-binding model. The height and width of resonators are $b = 28.4$ mm and $w = 7.5$ mm, respectively. The diameter of intracell hopping tubes is $d_1 = 1.5$ mm, while the intercell hopping tubes have a wider diameter $d_2 = 2.2$ mm. More details about the simulation and fabrication of PC are specified in Ref. [64]. The bulk dispersion of PC for $P_z$ mode (dipolar mode along the $z$-direction) of resonator along high-symmetry lines is depicted in Fig. 2(d), which overall agrees well with the result of lattice model. Along the Γ-M direction, a twofold degenerate nodal line emerges with a narrow frequency range of 6.05-6.09 kHz. The nodal line has a small slope because the chiral symmetry of real PC is broken slightly, which can be modified to lie a horizontal frequency by carefully designing the parameters of coupling tubes [68].

To experimentally demonstrate the nodal-line degeneracies, we fabricate a PC sample consisting of $20 \times 20 \times 6$ unit cells by employing 3D printing technology. The experiment setup is shown in Fig. 2(c). We place a sound source (cyan star) in the middle of the sample to excite bulk states, and probe the acoustic pressure in a horizontal plane slightly above the source. The measured pressure field exemplified at the nodal-line frequency (6.07 kHz) is depicted by the color map in Fig. 2(c). By further applying a 2D spatial Fourier transformation on the recorded sound signals, we can obtain projected bulk dispersions in the $k_x$-$k_y$ plane. Subsequently, we extract the projected bulk dispersions along the $k_x$ direction for different $k_y$, as shown in Fig. 2(e). The degenerate points for different $k_y$ can be observed clearly, which finally form the nodal lines in the $k_x$-$k_y$ plane. As a whole, the measured projected bulk dispersions (color maps) agree well with the simulated results [gray lines in Fig. 2(f)].

We then observe the drumhead surface states. As depicted in Fig. 3(a), we present a schematic of experimental setup for surface measurement. To excite surface states, a point-like sound source (cyan star) is inserted at the center of Surface I (a surface containing whole resonators, as elaborated in Ref. [64]). We scan the pressure field on the Surface I and obtain projected surface dispersions in the $k_x$-$k_y$ plane by applying the 2D Fourier transformation on the measured field. As shown in Fig. 3(b), we extract isofrequency contours for a series of frequencies. The color maps in the left panels represent the measured results, while the green and grey shadows in the right panels denote the simulated drumhead surface states and projected bulk states, respectively. For the panel of 6.10 kHz, drumhead surface states are observed clearly, which emerge at the region corresponding to that with $\pi$ Zak phase. While for the panels of other frequencies, only projected bulk states exist. We further extract the projected surface dispersions along two paths, as depicted in Fig. 3(c) for the $k_x$ direction with $k_y = 0$ and Fig. 3(d) for the $k_y$ direction with $k_x = \pi/a$, and confirm the flat feature of surface dispersion. Overall, the measured results are in good agreement with the numerical simulations (green lines).

Finally, we present the experimental observation of flat hinge states. To better observe the projected hinge dispersions along the $k_z$ direction, we fabricate another PC sample consisting of $6 \times 6 \times 21$ unit cells, as shown in Fig. 4(a). A sound source is placed at the middle of one hinge boundary to excite the hinge modes, as indicated by the cyan star in the inset of Fig. 4(c). We scan the pressure field on two adjacent surfaces of the concerned hinge and perform the Fourier transformation along the $z$ direction. Figure 4(b) shows such amplitude distribution of Fourier spectra in $x$-$y$-$k_z$ space at 6.10 kHz (the frequency of hinge state). It can be seen that the pressure field is localized at the boundaries between the surfaces, which visually indicate that the hinge states are localized for all $k_z$. By extracting the pressure field along the hinge and performing the Fourier transformation, we can obtain the measured projected hinge dispersions, as shown in Fig. 4(c). The color map represents the measured results, while the orange and grey lines denote the simulated hinge dispersion and projected bulk and surface dispersions, respectively. One can see that the hinge states are strongly excited and both the bulk and surface states are poorly excited, because the sound source is plugged into the hinge of sample. The measured hinge dispersion exhibits a good agreement with the simulated one. If a point source is inserted inside the sample, the bulk states are excited well, and the projected bulk dispersions with high densities of states can be captured clearly, as shown in Fig. 4(d). These results together evidence the existence of the hinge states.

In conclusion, we have theoretically proposed and experimentally realized an acoustic second-order NLSM. It hosts the gapless 2D surface states protected by Zak phase, verifying the conventional bulk-boundary correspondence. Meanwhile it also possesses the gapless 1D hinge states attributed to quadrupole moment, confirming the higher-order bulk-boundary correspondence. The nodal-line degeneracy, drumhead surface states, and flat hinge states are all observed in experiments. Our findings provide new avenues to multidimensionally manipulate the acoustic waves, which may have the potential for innovative applications in physical acoustics, including acoustic sensing and high-performance energy harvesting on surfaces and hinges [69,70]. In addition, our system can be extended to other artificial structures, such as photonic crystals [71,72] and mechanical metamaterials [73,74]. Besides our work, it is worth noting that two interesting works also focus on the 3D acoustic nodal-line semimetals recently, which possess Stiefel-Whitney topological charges and exhibit the PT-related hinge states with selected field distributions [75,76].

**Acknowledgements**

This work is supported by the National Key R&D Program of China (Grants No. 2022YFA1404900, No. 2022YFA1404500), the National Natural Science Foundation of China (Grants No. 11890701, No. 11974262, No. 11974120, No. 11974005, No. 12074128, No. 12004286, No. 12104347, No. 12222405, No. 12374409, No. 12374419), and Guangdong Basic and Applied Basic Research Foundation (Grants No. 2021B1515020086, No. 2022B1515020102).

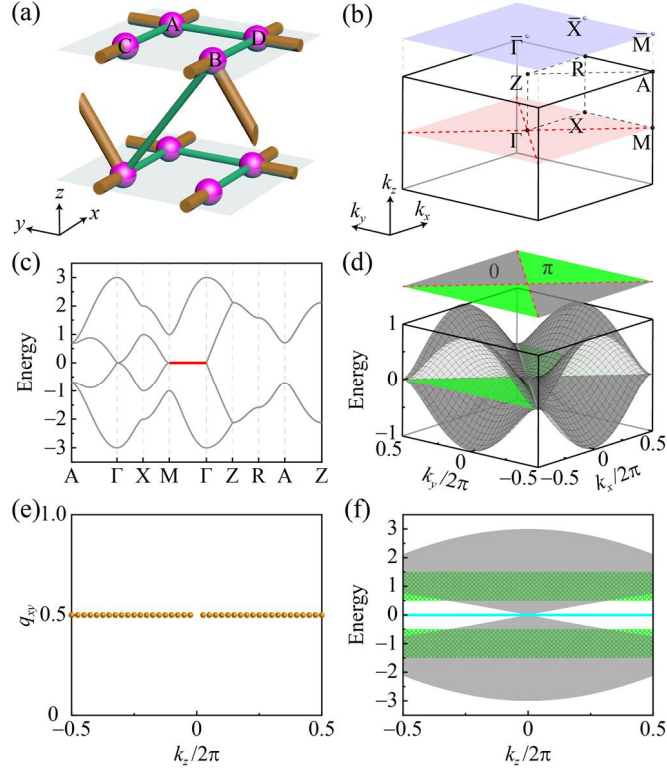

FIG. 1. Second-order NLSM for a simple cubic lattice model. (a) Schematic of tight-binding model. The unit cell consists of four sites A-D. Brown cylinders represent the hopping $t_0$, and green cylinders denote the hopping $t_1$. (b) The first BZ and corresponding surface BZ (blue plane). The red dotted lines label the nodal line. (c) Bulk band dispersions along the high-symmetry lines. (d) Upper panel: Zak phase distribution in the $k_x$-$k_y$ plane. The green (grey) region corresponds to the $\pi$ (0) Zak phase. Lower panel: projected surface dispersions. The green plane denotes the drumhead surface states and the grey area labels the bulk states. (e) Quadrupole moment distribution for the lowest two bulk bands as a function of $k_z$. (f) Projected hinge dispersions along the $k_z$ direction. The cyan lines represent the hinge states. Gray regions and green grids denote the projected bulk and surface states, respectively. The parameters are chosen as $t_1 = -0.5$ and $t_0 = -1$.

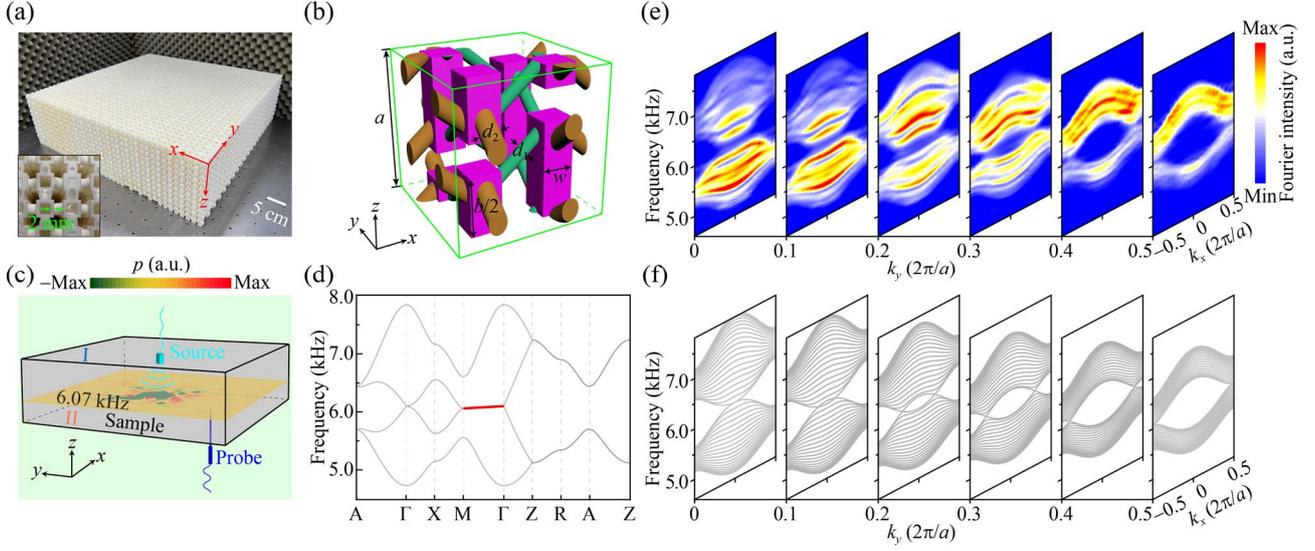

FIG. 2. Acoustic second-order NLSM and observation of the bulk nodal lines. (a) Photograph of the fabricated sample for bulk and surface measurements, with an enlarged top view of PC in the inset. (b) Unit cell of the PC. All the colored zones depict the area of air. Acoustic resonators (magenta cuboid) are connected by oblique coupling tubes (green and brown columns). (c) Experimental setup for measuring projected bulk dispersions. The measured pressure field $p$ within the sample at 6.07 kHz is depicted. (d) Bulk dispersions of PC along the high-symmetry lines. The red lines represent the nodal-line degeneracy. (e) and (f) Measured and simulated projected bulk dispersions along the $k_x$ direction for several $k_y$.

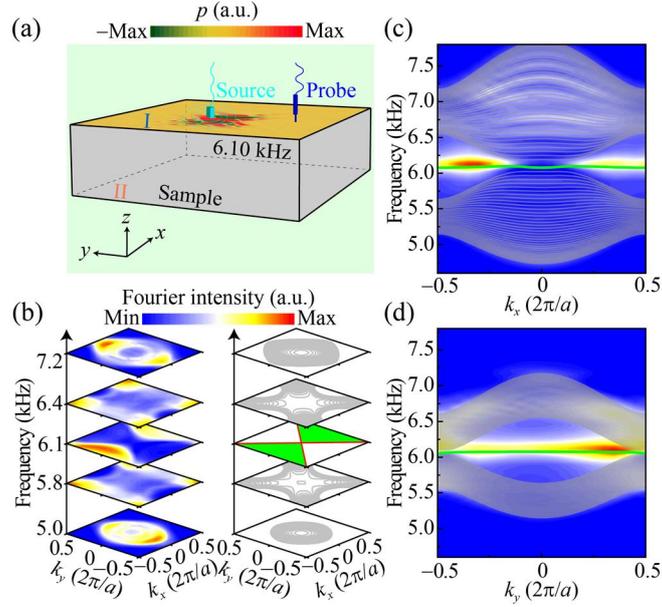

FIG. 3. Experimental observation of the drumhead surface states. (a) Experimental setup for measuring drumhead surface states. The measured pressure field $p$ on Surface I at 6.10 kHz is represented. (b) Measured isofrequency contours (left panels) by 2D Fourier transforming the surface fields for different frequencies, compared with the simulated results (right panels). The color maps, green shadows, and grey curves represent the measured results, simulated drumhead surface states and bulk states, respectively. (c) Measured and simulated projected surface dispersions along the $k_x$ direction for $k_y = 0$. (d) Measured and simulated surface dispersions along the $k_y$ direction for $k_x = \pi/a$. In (c) and (d), color maps denote the measured results, and green (grey) lines represent the simulated drumhead surface (projected bulk) states.

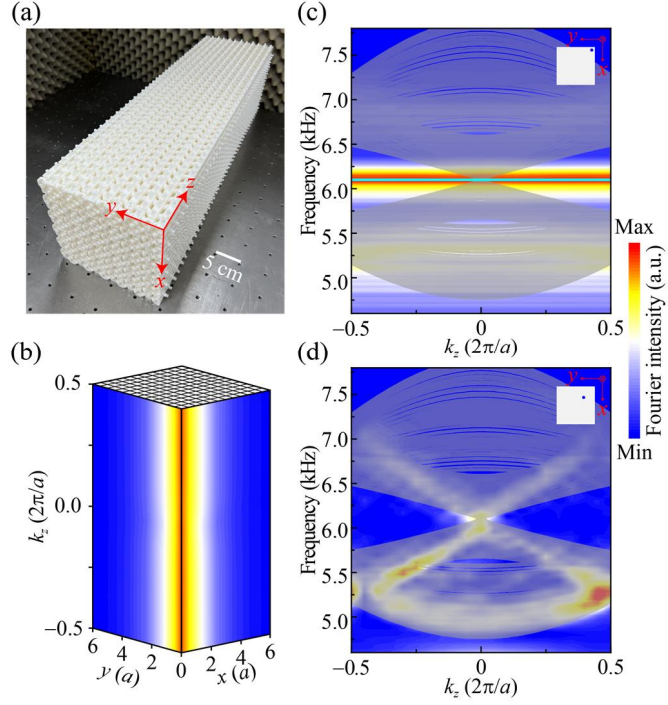

FIG. 4. Experimental observation of the hinge states. (a) Photograph of the fabricated sample for the hinge state measurement. (b) Measured pressure fields on two surfaces aside from a hinge along the $k_z$ direction at 6.10 kHz. The horizontal directions stand for the real space, while the vertical direction denotes the $k_z$ direction. (c) Measured and simulated projected hinge dispersions along the $k_z$ direction. Cyan (grey) lines represent simulated hinge (projected bulk) states. (d) Measured and simulated projected bulk dispersions along the $k_z$ direction. In (c) and (d), color maps represent measured results, and the cyan (grey) lines denote the simulated hinge (projected bulk) dispersions. The insets give the schematics of the measurement, where the bule circle labels the projections of the source and detector on the $x$-$y$ plane.